\documentstyle[12pt,psfig]{article}

\begin{document}
\baselineskip 0.7cm

\def\Slash#1{\ooalign{\hfil/\hfil\crcr$#1$}}

\newcommand{\gsim}{ \mathop{}_{\textstyle \sim}^{\textstyle >} }
\newcommand{\lsim}{ \mathop{}_{\textstyle \sim}^{\textstyle <} }
\newcommand{\vev}[1]{ \left\langle {#1} \right\rangle }
\newcommand{\lsp}{ \left ( }
\newcommand{\rsp}{ \right ) }
\newcommand{\lmp}{ \left \{ }
\newcommand{\rmp}{ \right \} }
\newcommand{\llp}{ \left [ }
\newcommand{\rlp}{ \right ] }
\newcommand{\labs}{ \left | }
\newcommand{\rabs}{ \right | }
\newcommand{\EV} { {\rm eV} }
\newcommand{\KEV}{ {\rm keV} }
\newcommand{\MEV}{ {\rm MeV} }
\newcommand{\GEV}{ {\rm GeV} }
\newcommand{\TEV}{ {\rm TeV} }
\newcommand{\YR}{ {\rm yr} }
\newcommand{\mgut}{M_{GUT}}
\newcommand{\mint}{M_{I}}
\newcommand{\mgra}{M_{3/2}}
\newcommand{\mll}{m_{\tilde{l}L}^{2}}
\newcommand{\mdr}{m_{\tilde{d}R}^{2}}
\newcommand{\mllXX}[1]{m_{\tilde{l}L , {#1}}^{2}}
\newcommand{\mdrXX}[1]{m_{\tilde{d}R , {#1}}^{2}}
\newcommand{\mgy}{m_{G1}}
\newcommand{\mgl}{m_{G2}}
\newcommand{\mgc}{m_{G3}}
\newcommand{\nuR}{\nu_{R}}
\newcommand{\slL}{\tilde{l}_{L}}
\newcommand{\slLi}{\tilde{l}_{Li}}
\newcommand{\sdR}{\tilde{d}_{R}}
\newcommand{\sdRi}{\tilde{d}_{Ri}}
\newcommand{\e}{{\rm e}}
\newcommand{\parml}{\partial_{\mu}}
\newcommand{\parmu}{\partial_{\mu}}
\newcommand{\Dmul}{D_{\mu}}
\newcommand{\Dmuu}{D^{\mu}}
\newcommand{\Gamu}{\Gamma^{u}}
\newcommand{\bsub}{\begin{subequations}}
\newcommand{\esub}{\end{subequations}}
\newcommand{\beqn}{\begin{eqnarray}}
\newcommand{\eeqn}{\end{eqnarray}}
\newcommand{\ra}{\rightarrow}
\newcommand{\wt}{\widetilde}
\newcommand{\btable}{\begin{table}[htbp]\begin{center}}
\newcommand{\etable}[1]{ \end{tabular}\caption{#1}\end{center}\end{table} }
\newcommand{\vf}{\vspace{0.5cm}}
\newcommand{\vt}{\vspace{0.2cm}}
\renewcommand{\thefootnote}{\fnsymbol{footnote}}
\setcounter{footnote}{1}

\makeatletter
%
%
%
%
%
\newtoks\@stequation

\def\subequations{\refstepcounter{equation}%
  \edef\@savedequation{\the\c@equation}%
  \@stequation=\expandafter{\theequation}
  \edef\@savedtheequation{\the\@stequation}
  \edef\oldtheequation{\theequation}%
  \setcounter{equation}{0}%
  \def\theequation{\oldtheequation\alph{equation}}}

\def\endsubequations{%
  \ifnum\c@equation < 2 \@warning{Only \the\c@equation\space subequation
    used in equation \@savedequation}\fi
  \setcounter{equation}{\@savedequation}%
  \@stequation=\expandafter{\@savedtheequation}%
  \edef\theequation{\the\@stequation}%
  \global\@ignoretrue}


\def\eqnarray{\stepcounter{equation}\let\@currentlabel\theequation
\global\@eqnswtrue\m@th
\global\@eqcnt\z@\tabskip\@centering\let\\\@eqncr
$$\halign to\displaywidth\bgroup\@eqnsel\hskip\@centering
     $\displaystyle\tabskip\z@{##}$&\global\@eqcnt\@ne
      \hfil$\;{##}\;$\hfil
     &\global\@eqcnt\tw@ $\displaystyle\tabskip\z@{##}$\hfil
   \tabskip\@centering&\llap{##}\tabskip\z@\cr}

\makeatother

\begin{titlepage}

\begin{flushright}
UT-976
\end{flushright}

\vskip 0.35cm
\begin{center}
{\Large \bf Sneutrino Factory}
\vskip 1.2cm
Yosuke Uehara

\vskip 0.4cm

{\it Department of Physics, University of Tokyo, 
         Tokyo 113-0033, Japan}
\vskip 1.5cm

\abstract{We argue that future $e^{+}e^{-}$ linear colliders can 
produce many sneutrinos if lepton-number violating couplings 
$\lambda_{1j1}$ exist and enough beam polarization is obtained. 
The terms $\lambda_{ijk} L_{i} L_{j} E_{k}^{c}$ are allowed in 
a discrete $Z_{3}$-symmetry which is used to forbid rapid proton decay,
and it is worthwhile to consider the possibility of the existence of
such terms and their resultant. 
We study the process $e^{+}e^{-} \ra \wt{\nu} \ra e^{+} e^{-}$ in 
detail, and show that if such resonance is not found, lepton-number violating
couplings $\lambda_{1j1}$, will be strongly constrained. If we assume
sneutrino mass $m_{\wt{\nu}}=500 \GEV$, beam polarization 
$P_{e^{-}}=0.9, \ P_{e^{+}}=0.6$, 
and integrated luminosity $L=100 {\rm fb}^{-1}$, non-observation of 
$\wt{\nu}$ resonance will lead to $\lambda_{1j1} \lsim 0.02$ 
(if all charginos, and neutralinos, are lighter than the sneutrino), or
$\lambda_{1j1} \lsim 0.003$ (if all charginos, and neutrainos,
are heavier than the sneutrino).}

\end{center}
\end{titlepage}

\renewcommand{\thefootnote}{\arabic{footnote}}
\setcounter{footnote}{0}

%
%
%
%

The Standard Model(SM) of particle physics explains many experimental
facts excellently, but it is not considered as the ultimate theory.
It is because Higgs sector of the SM receives divergent radiative
corrections, and a fine-tuning is needed to make the SM valid up 
to high energy scale. Supersymmetry(SUSY) is considered 
as the most promising symmetry to solve this fine-tuning problem. 
If we assume SUSY, the superpartner of the SM particles cancels the
divergent diagram and the Minimal Supersymmetric Standard Model(MSSM) 
can be valid up to high energy scale without any fine-tuning. 
Furthermore, the MSSM predicts the gauge coupling
unification at very high energy scale $E \sim 2 \times 10^{16} \GEV$, and thus
it is very motivated.

However the naive MSSM includes very dangerous terms such as
\begin{eqnarray}
W = \lambda_{ijk} L_{i} L_{j} E_{k}^{c} + \lambda^{'}_{ijk} L_{i} Q_{j} D_{k}^{c} + \lambda_{ijk}^{''} U_{i}^{c} D_{j}^{c} D_{k}^{c},
\label{RPVSUPERPOTENTIAL}
\end{eqnarray}
If these coefficients are ${\rm O}(1)$, we have too rapid proton decay
\cite{PROTONDECAY}. Thus, these coefficients must be extremely small.

The R-symmetry is a well-known symmetry to suppress the
superpotoential in Eq.(\ref{RPVSUPERPOTENTIAL}).
If one imposes R-parity invariance, superpotential in Eq. 
(\ref{RPVSUPERPOTENTIAL}) vanishes and no such a dangerous phenomenon occurs.
R-parity is defined as $R=(-1)^{(3B+L+2S)}$. Here $B$ is baryon number 
of the particle, $L$ is the lepton number of the particle, and
$S$ is the spin of the particle. 

However, there is another interesting symmetry to suppress 
the dangerous operators in Eq.(\ref{RPVSUPERPOTENTIAL}). 
A $Z_{3}$-symmetry forbids 
$\lambda_{ijk}^{''} U_{i}^{c} D_{j}^{c} D_{k}^{c}$-term completely 
and it is anomaly-free \cite{DISCRETESYMMETRY}.
Thus this $Z_{3}$-symmetry is interesting alternative to R-parity 
to suppress the dangerous rapid proton decay, since proton decay amplitude
is mediated by the terms 
proportional to the product $\lambda^{'} \lambda^{''}$.
The charge assignment of this $Z_{3}$ is shown in table \ref{Z3TAB}.

\btable
\begin{tabular}{|c|c|c|c|c|c|}
\hline
particle & $Q$ & $U^{c}$ & $D^{c}$ & $L$ & $E^{c}$ \\
\hline
charge & 1 & $\alpha^{2}$ & $\alpha$ & $\alpha^{2}$ & $\alpha^{2}$ \\
\hline
\etable{charge assignment under the discrete gauge symmetry
$Z_{3}$. \label{Z3TAB}}

If we assume this $Z_{3}$-symmetry, $\lambda_{ijk} L_{i} L_{j}
E_{k}^{c}$ and $\lambda_{ijk}^{'} L_{i} Q_{j} D_{k}^{c}$ is allowed,
and hence it is interesting to consider the phenomena which 
$\lambda_{ijk} L_{i} L_{j} E_{k}^{c}$ generates. In this paper,
we consider the sneutrio-resonant production caused by this term. 
We assume $\lambda_{ijk}^{'}= \lambda_{ijk}^{''}=0$ throughout this paper.

Constraints on the above lepton-number violating couplings 
$\lambda_{ijk} L_{i} L_{j} E_{k}^{c}$ can be obtained from 
experiments. We show that the non-observation of
superparticle in collider experiments can set upper bounds on couplings.
(``direct'' experiments)
On the other hand, we have effects of virtual superparticles 
at the quantum level, and can set upper bounds on couplings.
(``indirect'' experiments)

First, let us consider ``direct'' experiments.
LEP experiments searched for the lepton-number 
violating couplings \cite{L3,OPAL,ALEPH}. 
This direct search put strong constraints on $\lambda_{1j1}$, which we
will consider later, in the region $m_{\wt{\nu}} \lsim 200 \GEV$(LEP
maximum energy). 

Next, consider about ``indirect'' experiments \cite{INDIRECT}. 
They include the EDM of fermions \cite{EDM}, anomalous magnetic
moment of the muon \cite{MUONGMINUSTWO}, the decay width of the Z
\cite{LEPTOQUARKZ,LEPHADRONICZ,LEPTONUNIVERSALITYWZ} and of the W
\cite{LEPTONUNIVERSALITYWZ} bosons, the strength of four-fermion
interactions, with the subsequent production of lepton pairs at
hadron \cite{LEPTOQUARKANALYSIS,SUSYREPORT} and lepton colliders 
\cite{CONTACTINTLEP2}, rare processes such as $\mu \ra e \gamma$
\cite{MUEGAMMA}, $e - \mu - \tau$ universality 
\cite{EMUTAUUNIVERSALITY,LEPTONUNIVERSALITYWZ}.
(For an extensive discussion, see \cite{NEUTRALINO}.)

The existence of $\lambda$ coupling may 
wash-out the baryon-number asymmetry produced in the early universe 
\cite{BNONCONSERVATION}. This is because spharelon processes violate $B+L$,
and $\lambda$ couplings violate $L$, and thus the combination violates $B$.
Nevertheless, we can avoid this difficulty if 
the baryon-number asymmetry in the universe is created 
after the elecroweak phase transition, 
by the mechamism like Affleck-Dine baryogenesis 
\cite{AFFLECKDINE} or the electroweak baryogenesis \cite{EWBARYOGENESIS}.

Thus lepton-number violating couplings $\lambda_{ijk}$ 
are not severely constrained yet.
In this paper, we consider a ``direct'' experiment and the expected
constraints on the lepton-number violating couplings $\lambda_{1j1}$
in the future collider experiments.
The upper bounds of the lepton-number violating coupling $\lambda_{ijk}$ 
from many ``direct'' and ``indirect'' experiments are 
shown in table \ref{LAMBDABOUNDTABLE}.
\btable
\begin{tabular}{|c|c|c|}
\hline
$\lambda$ & upper bound & its origin \\
\hline
$\lambda_{121}$ & $<0.04$ & CC universarity \cite{LAMBDABOUND}. \\
\hline
$\lambda_{131}$ & $<0.10$ & tau leptonic decay \cite{LAMBDABOUND}. \\
\hline
\etable{The upper bound of lepton-number violation and their
 origin.\label{LAMBDABOUNDTABLE} This is model independent
constraint. Here we assume common SUSY scalar mass: $\wt{m}=100 \GEV$.}

The ``direct'' experiment we consider is the sneutrino single production
\begin{eqnarray}
e^{+} e^{-} \ra \wt{\nu} \ra e^{+} e^{-}.
\end{eqnarray}

It was previously studied in some articles 
\cite{INDIRECT,TESLATDR,RESONANTPRODUCTION,LEPTOQUARK}. 
But \cite{INDIRECT,TESLATDR,RESONANTPRODUCTION} did not consider the
interference between t-channel, \cite{TESLATDR,LEPTOQUARK} did not
calculate the decay width of sneutrino and they only assumed it,
and all of them did not take into account the effect of beam polarization. 

There are other articles which considered sneutrino resonant production.
\cite{SNEUTRINOBB} considered sneutrino s-channel production,
its decay into $b \bar{b}$, and their expected LEP signal.
\cite{MUONCOLLIDER1,MUONCOLLIDER2} considered muon collider discovery 
potential of $\lambda$ and $\lambda^{'}$ couplings.

The interference between resonant s-channel and t-channel
smears the peak of cross section, and so we must take it into account.
The decay width of sneutrino highly depends on the mass spectra of
SUSY particles, and it is very important for setting upper limit
on the lepton-number violating coupling from the experiments.
Finally, in order to distinguish this sneutrino $e^{+} e^{-}$
resonance from usual $\gamma/Z$ resonance tail in the case
that the sneutrino can decay into chargino and/or neutralino, 
we must use highly-polarized beam.

Sneutrino is produced as a s-channel resonance of $e^{+} e^{-}$ collision
if the lepton-number violating superpotential
\begin{eqnarray}
W = \lambda_{121} L_{1} L_{2} E_{1}^{c} + \lambda_{131} L_{1} L_{3} E_{1}^{c},
\end{eqnarray}
exists. We do not distinguish $\lambda_{121}$ and $\lambda_{131}$,
and simply write it as $\lambda$, hereafter.

Once produced, sfermion immediately decays into lighter particles.
In the following, we show the relevant Lagrangian for the decay of
sneutrino. Our convention is given in appendix \ref{SUSYCONVENTION}.

{\bf Lepton-number violation:}
\begin{eqnarray}
{\cal L} = \lambda \wt{\nu} e \bar{e}.
\label{RPVLAGRANGIAN}
\end{eqnarray}

{\bf Fermion-sfermion-chargino:}
\begin{eqnarray}
{\cal L} = &-& g \cos \phi_{R} (\wt{\nu}^{*} \wt{\chi}_{1}^{+} l_{L} + {\rm h.c.}) \nonumber \\
&+& g \epsilon_{R} \sin \phi_{R} (\wt{\nu}^{*} \wt{\chi}_{2}^{+} l_{L} + {\rm h.c.}).
\end{eqnarray}

{\bf Fermion-sfermion-neutralino:}
\begin{eqnarray}
{\cal L}&=& \sqrt{2} g_{Z} (-g_{NiL}^{*} \wt{f}_{L}^{*}
 \bar{\wt{\chi}}_{i}^{0} f_{L} + {\rm h.c.}), \\
g_{NiL} &=& \eta_{i} T_{3L} \cos \theta_{W} (O_{N})_{i2} + (Q-T_{3L}) \sin \theta_{W} (O_{N})_{i1}. \nonumber
\end{eqnarray}
There are other Lagrangian which causes the decay of sneutrino, 
but their decay modes are highly suppressed
mainly because of the multi-body phase space. So we do not write them.

The input parameters are described in table \ref{SUSYTABLE}, and the mass
spectra of neutralino and chargino are shown in table \ref{NCTABLE}.

\btable
\begin{tabular}{|c|c|c|}
\hline
parameter & mass(GeV) (case 1) & mass(GeV) (case 2) \\
\hline
$m_{\wt{\nu}}$ & 500 & 500 \\
\hline
$M_{1}$ & 100 & 600 \\
\hline
$M_{2}$ & 200 & 1200 \\
\hline
$\mu$ & 100 & 600 \\
\hline
\etable{SUSY input parameters. \label{SUSYTABLE}}

\btable
\begin{tabular}{|c|c|c|c|c|c|c|}
\hline
 & $\wt{\chi}_{1}^{0}$ & $\wt{\chi}_{2}^{0}$ & $\wt{\chi}_{3}^{0}$ &
 $\wt{\chi}_{4}^{0}$ & $\wt{\chi}_{1}^{-}$ & $\wt{\chi}_{2}^{-}$ \\ 
\hline
mass(GeV) (case 1) & 56.6 & 113 & 118 & 238 & 79 & 238 \\
\hline
mass(GeV) (case 2) & 563 & 602 & 530 & 1210 & 595 & 1210 \\
\hline
\etable{Neutralino and chargino mass spectrum \label{NCTABLE}.}

The decay modes
\bsub
\begin{eqnarray}
\wt{\nu} \ra \wt{\chi}_{1,2,3,4}^{0} + \nu , \\
\wt{\nu} \ra \wt{\chi}_{1,2}^{-} + l^{+} ,
\end{eqnarray}
\esub
are kinematically allowed in the case 1 and dominates the decay modes.  
In the case 2, such decay mode cannot be 
allowed and only the lepton-number violating decay
\begin{eqnarray}
\wt{\nu} \ra e^{+} e^{-},
\end{eqnarray}
dominates the decay. 
We will concentrate on this lepton-number violating decay mode. 

We calculate the decay width for the case 1 and the case 2.
The total decay width does not depend so much on $\lambda$ in the case 1 
because the decay mode is dominated by neutraino and chargino
decay channel,and typically it is:
\begin{eqnarray}
\Gamma_{\wt{\nu}} \sim 5.8 \ \GEV. \ {\rm (for \ the \ case \ 1)}
\end{eqnarray}

In the case 2, only the lepton-number violating decay is allowed and
it highly depend on $\lambda$. It is:
\begin{eqnarray}
\Gamma_{\wt{\nu}} \sim 0.0020 \ (\frac{\lambda}{0.01})^{2} \ \GEV. \ {\rm (for \ the \ case \ 2)}
\end{eqnarray}

Now we consider the production of sneutrino.
Since the sneutrino is spinless particle, beam polarization is extremely
useful for its production. It is well 
described in figure \ref{1POL9}, \ref{1POL2}, \ref{2POL9}, \ref{2POL2}. 
By using beam polarization, 
we can improve the production cross section of spinless
particle, $\wt{\nu}$. We cannot observe sharp peak in the case 1 if
beam polarization is not enough. We can observe very sharp peak
in the case 2 because in the case 2 sneutrino cannot decay into chargino
and neutraino, and thus total decay width is very small.
Here we selected the region $\cos \theta < \frac{1}{\sqrt{2}}$ in order to 
avoid the divergent t-channel photon exchange process. And we use CIRCE 
\cite{CIRCE} to take into account the effect of initial state radiation.
For CIRCE, we assume TESLA data as an input.

\begin{figure}[htbp]
\centerline{\psfig{file=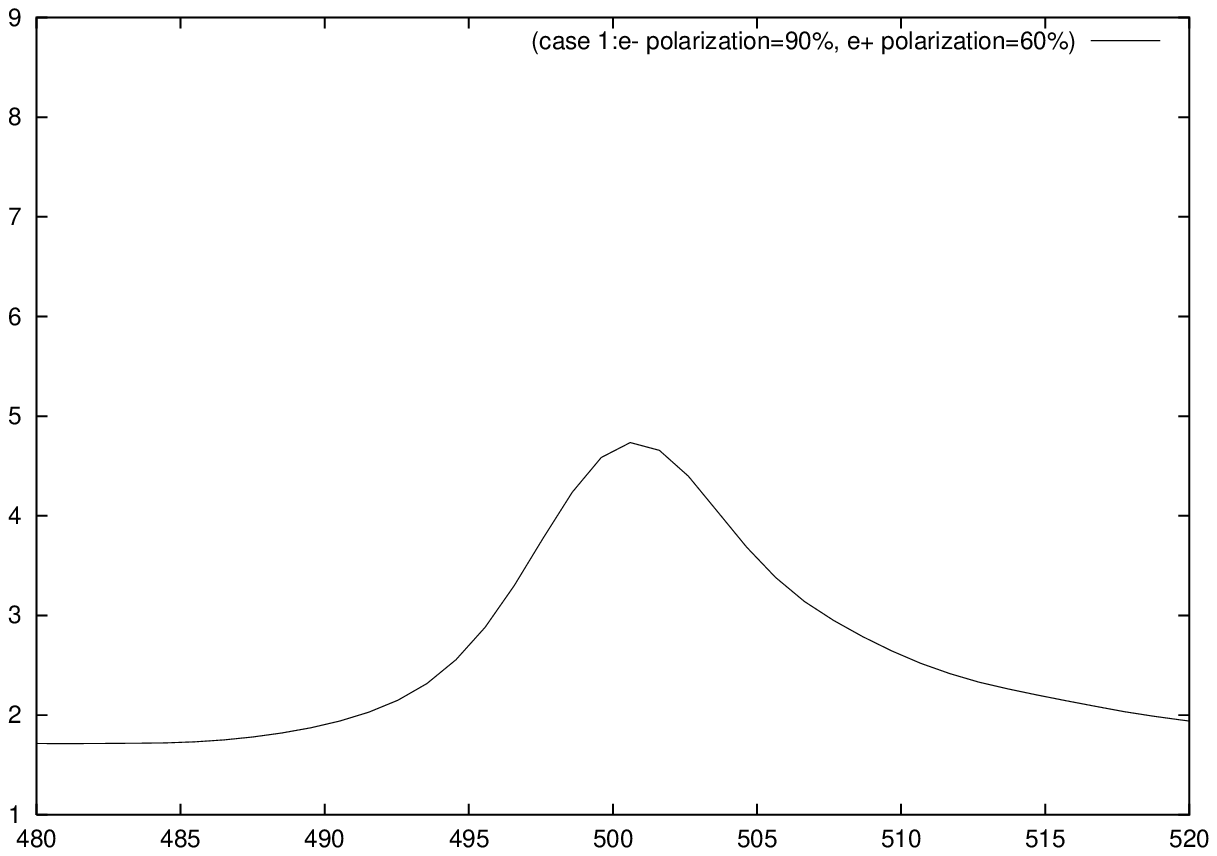,height=8cm}}
\begin{picture}(0,0)
\put(20,140){$\sigma ({\rm pb})$}  
\put(200,1){$\sqrt{s} \ (\GEV)$}
\end{picture}
\caption{Sneutrino resonance process $e^{+} e^{-} \ra \gamma/Z/\wt{\nu} \ra
 e^{+} e^{-}$ cross section in case 1. Here we assume $\lambda=0.05$ and
beam polarization $P_{e^{-}}=0.9, \ P_{e^{+}}=0.6$. We can observe the
peak of $\wt{\nu}$ around its mass, $m_{\wt{\nu}} = 500 \GEV$.\label{1POL9}}
\centerline{\psfig{file=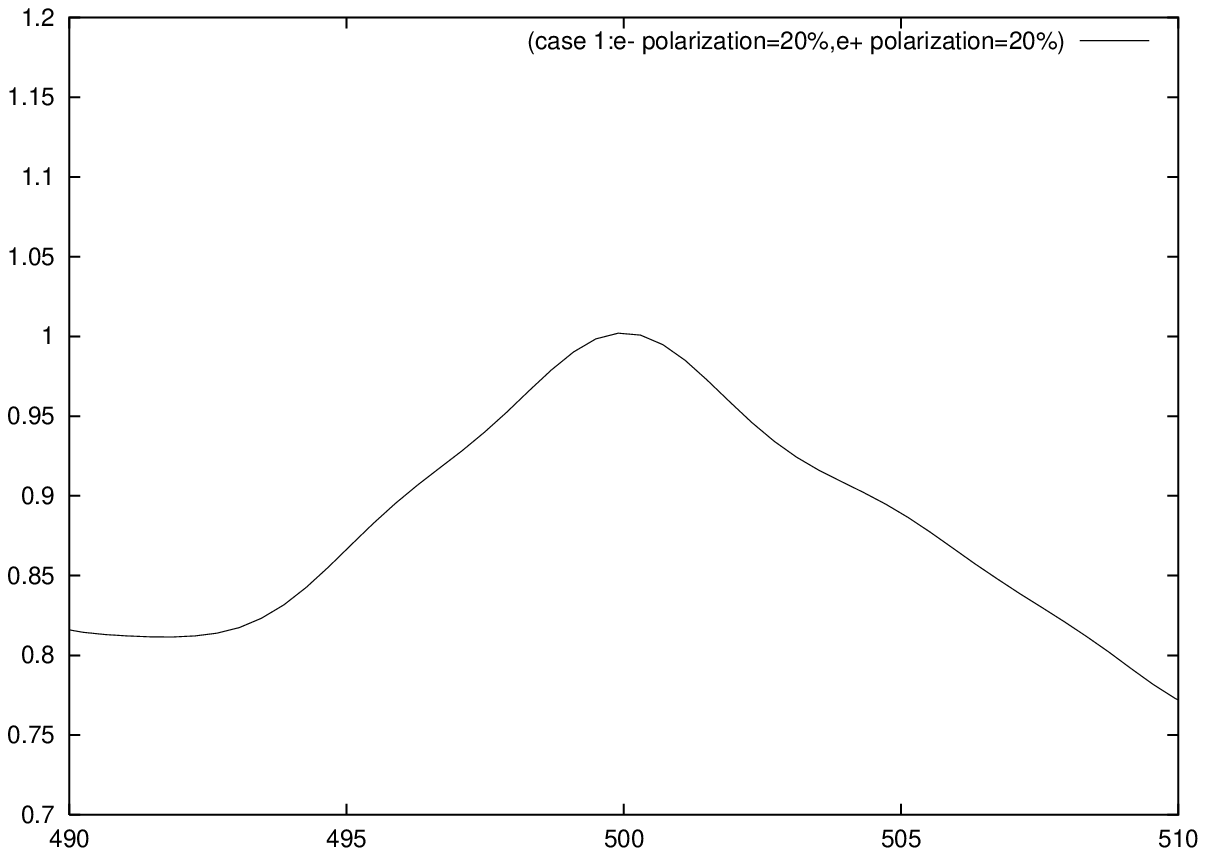,height=8cm}}
\begin{picture}(0,0)
\put(20,140){$\sigma ({\rm pb})$}  
\put(200,1){$\sqrt{s} \ (\GEV)$}
\end{picture}
\caption{Sneutrino resonance process $e^{+} e^{-} \ra \gamma/Z/\wt{\nu} \ra 
e^{+} e^{-}$ cross section in case 1. Here we assumed $\lambda=0.05$ and beam polarization $P_{e^{+}}=P_{e^{-}}=0.2$. We can see peak signal
slightly. \label{1POL2}}
\end{figure}

\begin{figure}[htbp]
\centerline{\psfig{file=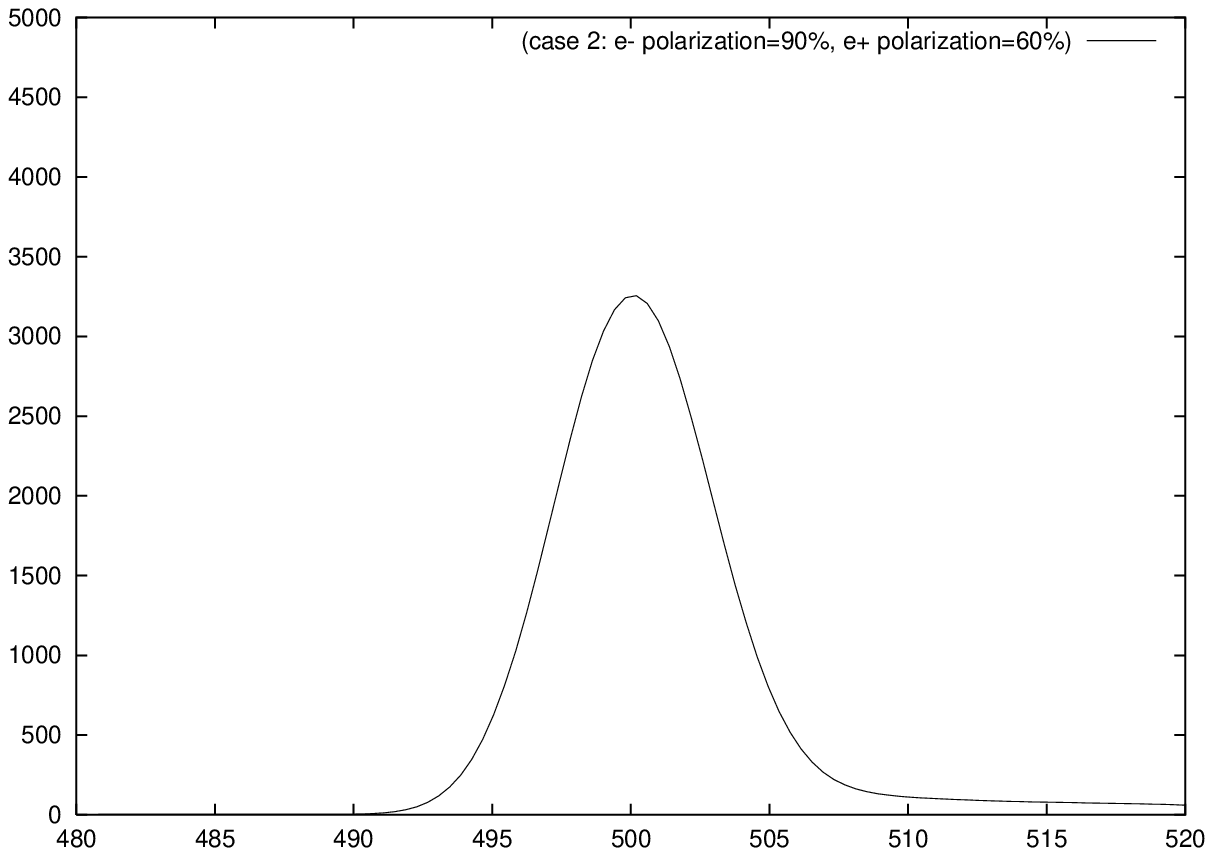,height=8cm}}
\begin{picture}(0,0)
\put(20,140){$\sigma ({\rm pb})$}  
\put(200,1){$\sqrt{s} \ (\GEV)$}
\end{picture}
\caption{Sneutrino resonance process $e^{+} e^{-} \ra \gamma/Z/\wt{\nu} \ra
 e^{+} e^{-}$ cross section in case 2. Here we assume $\lambda=0.05$ and
beam polarization $P_{e^{-}}=0.9, \ P_{e^{-}}=0.6$. We can observe the 
extremelly sharp peak of $\wt{\nu}$ around its mass, $m_{\wt{\nu}} = 500 \GEV$.
The sharpness comes from the small decay width of sneutrino. In case
2, sneutrino cannot decay into chargino and neutralinos. \label{2POL9}}
\centerline{\psfig{file=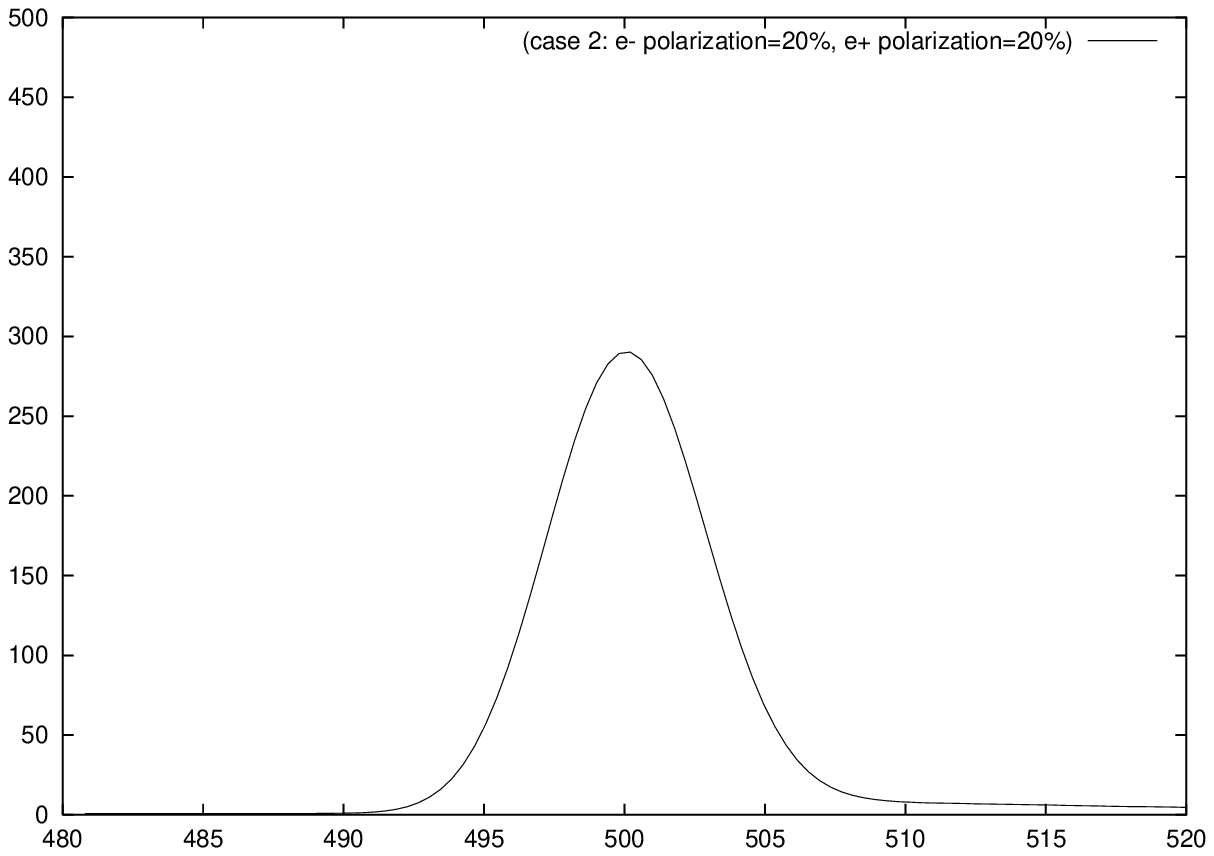,height=8cm}}
\begin{picture}(0,0)
\put(20,140){$\sigma ({\rm pb})$}  
\put(200,1){$\sqrt{s} \ (\GEV)$}
\end{picture}
\caption{Sneutrino resonance process $e^{+} e^{-} \ra \gamma/Z/\wt{\nu} \ra 
e^{+} e^{-}$ cross section in case 2. Here we assume $\lambda=0.05$ and
beam polarization $P_{e^{+}}=P_{e^{-}}=0.2$. Here again we can observe the
 peak of $\wt{\nu}$. This is because of the smallness of sneutrino decay width..\label{2POL2}}
\end{figure}

So with the help of beam polarization, we may be able to
find the s-channel resonance of sneutrino even in the case 1. 
If sneutrino is lighter than charginos and neutrainos, we may be 
able to see resonance without beam polarization.

If such resonance cannot be found, lepton-number 
violating coupling $\lambda$ will
be severely constrained.Here we assume 
integrated luminosity $100 {\rm fb}^{-1}$.
Half of the experiment is done with beam polarization 
$P_{e^{-}}=0.9, \ P_{e^{+}}=0.6$ and other half of the experiment
is done with beam polarization $P_{e^{-}}=0.9, \ P_{e^{+}}=-0.6$.
Then we can obtain the upper limit on $\lambda$ if we cannot
see $e^{+} e^{-} \ra \wt{\nu} \ra e^{+} e^{-}$ resonance.
For a sneutrino mass of $500 \GEV$,
\bsub
\begin{eqnarray}
\lambda &\lsim& 0.021 \ ({\rm for \ the \ case \ 1}), \\
\lambda &\lsim& 0.0030 \ ({\rm for \ the \ case \ 2}).
\end{eqnarray}
\esub

To summarize, we consider the process 
$e^{+} e^{-} \ra \wt{\nu} \ra e^{+} e^{-}$ in the 
lepton-number braking models. 
Such processes are allowed in anomaly-free $Z_{3}$-symmetry
which is imposed to protect the rapid proton decay
 \cite{DISCRETESYMMETRY} instead of the $Z_{2}$-symmetry, ``R-parity''.
We observe that using beam polarization is extremely useful 
to see the resonant peak of sneutrino in $e^{+} e^{-}$ system. 
Non-observation of such resonance will lead 
to a strong constraint on the lepton-number violating parameter $\lambda$.

\noindent
{\bf Acknowledgement}

We thank T.Yanagida and F.Borzumati for useful discussions.

\newpage

\appendix

\section{Convention of Supersymmetric Interaction}
\label{SUSYCONVENTION}

I followed the convention of ``Supersymmetric Standard Model
for Collider Physicists'' \cite{HIKASANOTE}.

\subsection{Chargino Sector}
\label{CHARGINO}

The chargino mass matrix $M_{C}$ can be written as:
\begin{eqnarray}
M_{C}=
\pmatrix{
M_{2} & \sqrt{2} m_{W} \cos \beta \cr
\sqrt{2} m_{W} \sin \beta & \mu},
\end{eqnarray}
here we asume $CP$ invariance and take all the entries in $M_{C}$ real.
then $M_{C}$ can be diagonalized by two orthogonal matrices:

\begin{eqnarray}
O_{R} M_{C} O_{L}^{\rm T} = {\rm diag},
\end{eqnarray}
where
\begin{eqnarray}
O_{L,R}=\pmatrix
{\cos \phi_{L,R} & \sin \phi_{L,R} \cr
-\sin \phi_{L,R} & \cos \phi_{L,R}}.
\end{eqnarray}

This diagonalization may leave negative eigenvalue in the diagonal entry.
So we make every entry positive. we first select the angle $\phi_{L}$
in the range $0 < \phi_{L} < \pi$, then $m_{\chi_{1}}^{-}$ can be made
positive if we allow the full range for $\phi_{R}$: $- \pi < \phi_{R} <
\pi$. Now $m_{\chi_{2}^{-}}$ can be either positive or negative
depending on sgn($M_{2} \mu - m_{W}^{2} \sin 2 \beta$). We redefine
$O_{R}$ as to allow for det$O_{R} =-1$:
\begin{eqnarray}
O_{R}=
\pmatrix
{\cos \phi_{R} & \sin \phi_{R} \cr
-\epsilon_{R} \sin \phi_{R} & \epsilon_{R} \cos \phi_{R}},
\end{eqnarray}
where $\epsilon_{R} = {\rm sgn}(M_{2} \mu - m_{W}^{2} \sin 2 \beta)$.

\subsection{Neutraino Sector}

The neutralino mass matrix can be written as:
\begin{footnotesize}
\begin{eqnarray}
M_{N}=\pmatrix{
M_{1} & 0 & -m_{Z} \sin \theta_{W} \cos \beta & m_{Z} \sin \theta_{W} \sin \beta \cr
0 & M_{2} & m_{Z} \cos \theta_{W} \cos \beta & -m_{Z} \cos \theta_{W} \sin \beta \cr
-m_{Z} \sin \theta_{W} \cos \beta & m_{Z} \cos \theta_{W} \cos \beta & 0 & -\mu \cr
m_{Z} \sin \theta_{W} \sin \beta & -m_{Z} \cos \theta_{W} \sin \beta & -\mu & 0 \cr}
\end{eqnarray}
\end{footnotesize}

Hereafter we assume that all the entries in $M_{N}$ are real. In this
case $M_{N}$ can be diagonized by an orthogonal matrix $O_{N}$:
\begin{eqnarray}
O_{N} M_{N} O_{N}^{\rm T} = {\rm real \ diagonal}. 
\label{NEUTRALINODIAG}
\end{eqnarray}
However this procedure may leave some of the eigenvalues negative.
The negative eigenvalues cannot be treated by introduging an extra sign
factor, as in Section \ref{CHARGINO}, because the left and right
neutralino fields are not independent. It is necessary to multiply the
negative mass eigenstates by a factor of $i$. In practice, this is done
by introducing an extra phase matrix in (\ref{NEUTRALINODIAG}).
\begin{eqnarray}
U_{N}^{*} M_{N} U_{N}^{\dagger} = {\rm positive \ diagonal},
\end{eqnarray}
where $U_{N}= \Phi_{N} O_{N}$ with $O_{N}$ given above, and $\Phi_{N}$
is a diagonal phase matrix $(\Phi_{N})_{ij}=\delta_{ij} \eta_{i}$ with
\begin{eqnarray}
\eta_{i}= \left\{\begin{array}{c} 1 \ \ \ {\rm if \ the \ mass \ eigenvalue \ is\  positive,} \\ i \ \ \ {\rm  if \ the \ mass \ eigenvalue \ is \ negative.} 
\end{array}\right.
\end{eqnarray}

%
%
\newcommand{\Journal}[4]{{\sl #1} {\bf #2} {(#3)} {#4}}
\newcommand{\PL}{\sl Phys. Lett.}
\newcommand{\PR}{\sl Phys. Rev.}
\newcommand{\PRL}{\sl Phys. Rev. Lett.}
\newcommand{\NP}{\sl Nucl. Phys.}
\newcommand{\ZP}{\sl Z. Phys.}
\newcommand{\PTP}{\sl Prog. Theor. Phys.}
\newcommand{\NC}{\sl Nuovo Cimento}
\newcommand{\PAN}{\sl Phys.Atom.Nucl.}

\end{document}